\documentclass[fleqn,usenatbib]{mnras}

\usepackage{newtxtext,newtxmath}

\usepackage[T1]{fontenc}

\usepackage{graphicx}	
\usepackage{amsmath}	
\usepackage{gensymb}
\usepackage{hyperref}

\title[MAXI J1820+070 Ejecta]{The Varying Kinematics of Multiple Ejecta from the Black Hole X-ray Binary MAXI J1820+070}

\author[C. M. Wood et al.]{C. M. Wood,$^{1}$\thanks{E-mail: callan.m.wood@student.curtin.edu.au}
J. C. A. Miller-Jones,$^{2}$
J. Homan,$^{3,4}$
J. S. Bright,$^{5}$
S. E. Motta,$^{5,6}$
R. P. Fender,$^{5,7}$
\newauthor S. Markoff,$^{8,9}$
T. M. Belloni,$^{6}$
E. G. K\"{o}rding,$^{10}$
D. Maitra,$^{11}$
S. Migliari,$^{12,13}$
D. M. Russell,$^{14}$
\newauthor T. D. Russell,$^{8,15}$
C. L. Sarazin,$^{16}$
R. Soria,$^{17,18}$
A. J. Tetarenko$^{19}$ and
V. Tudose$^{20}$
\\
$^{1}$Curtin University Department of Physics and Astronomy, Perth, WA 6845, Australia\\
$^{2}$International Centre for Radio Astronomy Research, Curtin University, GPO Box U1987, Perth, WA 6845, Australia\\
$^{3}$Eureka Scientific, Inc., 2452 Delmer Street, Oakland, CA 94602, USA\\
$^{4}$SRON Netherlands Institute for Space Research, Sorbonnelaan 2, NL-3584 CA Utrecht, The Netherlands\\
$^{5}$Astrophysics, Department of Physics, University of Oxford, Keble Road, Oxford, OX1 3RH, UK\\
$^{6}$INAF–Osservatorio Astronomico di Brera, Via E. Bianchi 46, I-23807 Merate (LC), Italy\\
$^{7}$Department of Astronomy, University of Cape Town, Private Bag X3, Rondebosch 7701, South Africa\\
$^{8}$Anton Pannekoek Institute for Astronomy, University of Amsterdam, Science Park 904, NL-1098 XH Amsterdam, The Netherlands\\
$^{9}$Gravitation \& Astroparticle Physics Amsterdam Institute, University of Amsterdam, NL-1098 XH Amsterdam, The Netherlands\\
$^{10}$Department of Astrophysics/IMAPP, Radboud University, PO 9010, NL- 6500 GL, Nijmegen, The Netherlands\\
$^{11}$Department of Physics and Astronomy, Wheaton College, Norton, MA 02766, USA\\
$^{12}$XMM–Newton Science Operations Centre, ESAC/ESA, Camino Bajo del Castillo s/n, Urb. Villafranca del Castillo, E-28691 Villanueva de la Ca\~{n}ada,\\ Madrid, Spain\\
$^{13}$Institute of Cosmos Sciences, University of Barcelona, Mart\'{i} i Franqu\`{e}s 1, E-08028 Barcelona, Spain\\
$^{14}$Center for Astro, Particle and Planetary Physics, New York University Abu Dhabi, PO Box 129188, Abu Dhabi, UAE\\
$^{15}$INAF, Istituto di Astrofisica Spaziale e Fisica Cosmica, Via U. La Malfa 153, I-90146 Palermo, Italy\\
$^{16}$Department of Astronomy, University of Virginia, 530 McCormick Road, Charlottesville, VA 22904-4325, USA\\
$^{17}$College of Astronomy and Space Sciences, University of the Chinese Academy of Sciences, Beijing 100049, China\\
$^{18}$Sydney Institute for Astronomy, School of Physics A28, The University of Sydney, Sydney, NSW 2006, Australia\\
$^{19}$East Asian Observatory, 660 N. A'oh\={o}k\={u} Place, University Park, Hilo, HI 96720, USA\\
$^{20}$Institute for Space Sciences, Atomistilor 409, PO Box MG-23, 077125 Bucharest-Magurele, Romania
}

\date{Accepted XXX. Received YYY; in original form ZZZ}

\pubyear{2020}

\begin{document}
\label{firstpage}
\pagerange{\pageref{firstpage}--\pageref{lastpage}}
\maketitle

\begin{abstract}
    During a 2018 outburst, the black hole X-ray binary MAXI J1820+070 was comprehensively monitored at multiple wavelengths as it underwent a hard to soft state transition. During this transition a rapid evolution in X-ray timing properties and a short-lived radio flare were observed, both of which were linked to the launching of bi-polar, long-lived relativistic ejecta. We provide detailed analysis of two Very Long Baseline Array observations, using both time binning and a new dynamic phase centre tracking technique to mitigate the effects of smearing when observing fast-moving ejecta at high angular resolution. We identify a second, earlier ejection, with a lower proper motion of $18.0\pm1.1$ mas day$^{-1}$. This new jet knot was ejected $4\pm1$ hours before the beginning of the rise of the radio flare, and $2\pm1$ hours before a switch from type-C to type-B X-ray quasi-periodic oscillations (QPOs). We show that this jet was ejected over a period of $\sim6$ hours and thus its ejection was contemporaneous with the QPO transition. Our new technique locates the original, faster ejection in an observation in which it was previously undetected. With this detection we revised the fits to the proper motions of the ejecta and calculated a jet inclination angle of $(64\pm5)\degree$, and jet velocities of $0.97_{-0.09}^{+0.03}c$ for the fast-moving ejecta ($\Gamma>2.1$) and $(0.30\pm0.05)c$ for the newly-identified slow-moving ejection ($\Gamma=1.05\pm0.02$). We show that the approaching slow-moving component is predominantly responsible for the radio flare, and is likely linked to the switch from type-C to type-B QPOs, while no definitive signature of ejection was identified for the fast-moving ejecta.
\end{abstract}

\begin{keywords}
stars: black holes -- X-rays: binaries -- stars: individual: MAXI J1820+070 -- stars: jets -- accretion, accretion discs -- techniques: high angular resolution
\end{keywords}



\section{Introduction}
    Low mass X-ray binary (LMXB) systems consist of either a stellar-mass black hole or a neutron star accreting mass from a low-mass binary companion star. Observations of such objects at X-ray and radio frequencies have identified two canonical accretion states, the hard state and the soft state. During outbursts, LMXBs typically undergo transitions between the hard and soft states via intermediate states \citep{homan2005evolution}. One of the defining characteristics of the hard state is the presence of strong, continuous jets, which are not present in the soft state. Discrete, transient jets are seen at the transition from the hard to the soft state, but not during the reverse transition \citep{fender2004towards}. During the transition from the hard to the soft state via intermediate states, a rapid evolution of the X-ray timing properties is seen. At the beginning of the transition, strong low-frequency type-C quasi-periodic oscillations (QPOs) are usually present, but are later often replaced by type-B QPOs \citep[see][for a discussion of low frequency QPOs]{Ingram2020}. The presence of type-B QPOs is thought to be linked to the launching of transient jets \citep[e.g.][]{fender2009jets,millerjones2012,Russell_2019}. 
    
    Accretion states and their relationship with the formation of relativistic jets have long been studied to understand the dynamics of jet launching events \citep{fender2004towards}. One focus of such studies has been attempting to confirm the causal connection between changes in the accretion flow and the launching of transient ejecta at the state transition. While suggestions have been made that particular spectral or timing signatures correspond to the moment of jet launching \citep[e.g.][]{fender2009jets,millerjones2012}, the relative sparsity of high angular resolution coverage and the uncertainty in the derived jet ejection times has meant that we still do not have a definitive signature of the changes in the accretion flow that lead to the launching of the transient jets.

    MAXI J1820+070/ASASSN-18ey (hereafter J1820) was first discovered at optical wavelengths on 7th March 2018 by the All-Sky Automated Survey for SuperNovae \citep[ASAS-SN;][]{tucker2018asassn}, and identified as a new X-ray binary system following a detection at X-ray wavelengths on 11th March by the Monitor of All-Sky X-Ray Image (MAXI) \citep{kawamuro2018maxi}.  It has since been dynamically confirmed to host a black hole \citep{torres2019dynamical}, and radio parallax measurements in the hard state have determined its distance to be $2.96 \pm 0.33$ kpc \citep{atri2020radio}, consistent with the value of $2.66_{-0.52}^{+0.85}$ kpc from Gaia Early Data Release 3 \citep{GAIA2020}, after applying the position-dependent zero point correction \citep{lindergren2020} and using the \citet{atri2019} prior.
    
    J1820 was in the hard state between its discovery in March 2018 and July 2018, when it underwent a hard-to-soft state transition \citep{Homan2018a,tetarenko2018Atel}. J1820 stayed in the soft state until the beginning of October 2018, when it returned to the hard state \citep{Homan2018b}. During its 2018 outburst J1820 was observed extensively at multiple different wavelengths \citep[e.g.][]{Shidatsu_2019,homan2020rapid,wang2020,kosenkov2020,buisson2021}. The transition between the hard state and the soft state, where J1820 went through the intermediate states, occurred between MJD 58303.5 and 58310.7  \citep{Shidatsu_2019}. X-ray coverage with the Neutron Star Interior Composition Explorer (\textit{NICER}) revealed rapid changes in the X-ray variability properties of J1820 during the hard-to-soft state transition. A switch from type-C QPOs to type-B QPOs was seen, along with a small flare in the 7--12 keV band \citep{homan2020rapid}. Shortly following this change in the X-ray variability properties, \citet{bright2020} reported on a short-lived ($\approx$12 hrs) radio flare beginning at MJD $58305.773\pm0.006$, which was detected using the Arcminute Microkelvin Imager-Large Array (AMI-LA). \citet{homan2020rapid} linked the change in X-ray variability properties and the radio flare to the ejection of two long-lived, apparently superluminal ejecta monitored by \citet{bright2020} as they travelled in opposite directions away from the core.  
    
    Radio observations of these ejecta with the Multi-Element Radio Linked Interferometer Network (eMERLIN), Meer Karoo Array Telescope (MeerKAT), the Karl G. Jansky Very Large Array (VLA) and the Very Long Baseline Array (VLBA) spanning a period of over 200\,d show the ejecta travelling out to angular separations of several arcseconds \citep[$\sim3\times10^4$ AU projected on the plane of the sky;][]{bright2020}. The approaching jet component was travelling to the south and the corresponding receding component to the north. Following these radio detections, the \textit{Chandra} X-ray telescope was triggered to observe the downstream re-brightening of the jets as they decelerated upon colliding with a denser region of the interstellar medium, creating external shocks. Using the radio data of \citet{bright2020} combined with the \textit{Chandra} observations, \citet{espinasse2020relativistic} fit the proper motions of these ejecta with a constant deceleration model. They found initial proper motions of $35.9\pm0.5$ and $93.3\pm0.6$ mas day$^{-1}$ for the north and south components, respectively, with accelerations of $-0.045 \pm 0.004$ and $-0.34 \pm 0.01$ mas day$^{-2}$, respectively, and an inferred ejection date of MJD $58305.97 \pm 0.07$. 

    High angular resolution imaging of ejecta travelling with such large proper motions can result in smearing of the image as components travel significant fractions of a resolution element during the observation. This violates a fundamental assumption of very long baseline interferometry (VLBI), that a source remains unchanged over the course of an observation. Multiple approaches have been used to image dynamical systems, from a relatively straightforward time binning approach \citep[e.g.][]{Fomalont_2001,V404_Cygni} to the more sophisticated dynamical imaging procedure devised by the Event Horizon Telescope consortium \citep{Dynamical_EHT}.
    
    Here we describe a new technique for reducing smearing in images of fast moving ejecta. We use this technique to detect the approaching fast-moving relativistic ejection described by \citet{bright2020} and \citet{espinasse2020relativistic} in a VLBA observation of J1820, in which this component was previously undetected. We provide a refined analysis of two VLBA observations, and via time binning, identify the previously-detected VLBI jet knot as a separate, slow moving component distinct from the fast moving ejecta tracked by \citet{bright2020}.  With this information, we revise the fits to the proper motions of the fast moving ejecta, and consider the implications for the physical parameters of the jet and the coupling between changes in the accretion flow and jet ejection.
    
\section{Methods}
    \subsection{Observations and Calibration}
       
    Following the initial detection of the outburst in the X-ray \citep{kawamuro2018maxi} and optical \citep{tucker2018asassn} bands, we observed J1820 with the VLBA over multiple epochs between 2018 March 16 and December 22, under project code BM467.  We took several epochs of astrometric data in the hard states at the beginning and end of the outburst, as detailed by \citet{atri2020radio}.  However, in this work we focus only on the data taken during and immediately after the hard-to-soft state transition in 2018 July.  We observed on twelve days between July 7 and 25, at a frequency of either 4.98 or 15.26\,GHz, depending on the weather and the source behaviour. Table~\ref{tab:observations} lists the parameters of the observations. 
        \begin{table}
            \centering
            \caption{VLBA observation log.}
            \label{tab:observations}
            \begin{tabular}{|c|c|c|c|c|}
            \hline
            Epoch          & Date & MJD                 & Time          & Frequency \\ 
            & (d/m/y) & & (UTC) & (GHz)\\ \hline \hline
            1              & 07/07/2018 &  $58306.22\pm0.02$ & 04:51                  - 05:38   & 15.26            \\ 
            2              & 07/07/2018 & $58306.37\pm0.02$ & 08:25                  - 09:08    &  15.26          \\ 
            3              & 08/07/2018 & $58307.27\pm0.04$ & 05:39                  - 07:22    &  15.26          \\ 
            4              & 09/07/2018 & $58308.39\pm0.02$ & 08:51                  - 09:38    &  \phantom{1}4.98          \\
            5              & 10/07/2018 & $58309.18\pm0.02$ & 03:51                  - 04:38    &  \phantom{1}4.98          \\ 
            6              & 10/07/2018 & $58309.39\pm0.02$ & 08:51                  - 09:38    &  \phantom{1}4.98          \\ 
            7              & 11/07/2018 & $58310.18\pm0.02$ & 04:51                  - 04:37    &  \phantom{1}4.98          \\ 
            8              & 11/07/2018 & $58310.35\pm0.02$ & 08:09                  - 08:52    &  \phantom{1}4.98          \\ 
            9              & 13/07/2018 & $58312.19\pm0.02$ & 04:06                  - 04:52    &  \phantom{1}4.98          \\  
            10             & 13/07/2018 & $58312.34\pm0.02$ & 07:54                  - 08:37    &  \phantom{1}4.98          \\
            11             & 14/07/2018 & $58313.25\pm0.02$ & 05:36                  - 06:22    &  \phantom{1}4.98          \\
            12             & 16/07/2018 & $58315.32\pm0.02$ & 05:36                  - 06:22    &  \phantom{1}4.98          \\
            13             & 18/07/2018 & $58317.30\pm0.02$ & 06:51                  - 07:37    &  15.26          \\
            14             & 20/07/2018 & $58319.23\pm0.02$ & 05:06                  - 05:52    &  \phantom{1}4.98          \\
            15             & 22/07/2018 & $58321.29\pm0.04$ & 06:09                  - 07:53    &  15.26          \\
            16             & 25/07/2018 & $58324.19\pm0.02$ & 03:39                  - 05:22    &  \phantom{1}4.98          \\
            \hline
            
            \end{tabular}

        \end{table}

    We observed with a recording rate of 2048\,Mbps, yielding a bandwidth of 256\,MHz per polarization, split into eight 32-MHz intermediate-frequency (IF) pairs.  We used ICRF J180024.7+384830 \citep[][hereafter J1800+3848]{Ma1998} as a fringe finder, ICRF J181333.4+061542 \citep[][hereafter J1813+0615]{beasley2002} as a phase reference calibrator, and RFC J1821+0549 as a check source.  The data were correlated using the DiFX software correlator \citep{Deller2011}, and calibrated according to standard procedures within the Astronomical Image Processing System \citep[\textsc{AIPS}, version 31DEC17;][]{wells1985nrao,Greisen}.  Following a priori corrections to the Earth Orientation Parameters at the time of correlation, we corrected for the ionospheric Faraday rotation and dispersive delay using Total Electron Content maps.  We then performed a priori corrections for digital sampler offsets and the changing parallactic angles of the feeds, before calibrating the amplitudes using the recorded system temperature values.  We used the bright fringe finder source J1800+3848 to determine the instrumental frequency response, and to correct instrumental phases, delays and rates. Finally, we performed several rounds of self calibration on the phase reference calibrator J1813+0615 to make the best possible model, which was used to compute time-varying phase, delay and rate solutions that were interpolated to the target source.
        
        \subsection{Imaging}
        Imaging was performed using \textsc{AIPS}, implementing the CLEAN algorithm \citep{hogbom1974clean}. Only the first two epochs, both taken on July 7th but separated by $\sim 3$\,hrs, yielded detections of J1820, likely due to a combination of the low brightness and adiabatic expansion of the jet ejecta.  We therefore focus on these epochs for the rest of this work.
        
        The fast moving ejecta described by \citet{bright2020} and \citet{espinasse2020relativistic} travelled across the synthesized beam of the VLBA observations in as little as 10 minutes, which is shorter than the length of the observations. As a result, these components are significantly smeared out in our VLBA images. To mitigate this, we adopted two different approaches.
        \subsubsection{Time Binning}\label{sec:timebinning}
            We imaged the first observation (epoch 1) in full, then split it into 5 time bins of length $\approx 9$ mins, each of which we subsequently imaged. We chose the size of the time bins to be as small as possible such that a distinct jet component could still be detected in each time bin. The second observation (epoch 2) could not be time binned due to a low signal to noise ratio (SNR) so we treated it as a single time bin. This time binning was performed in order to determine the proper motions of components seen in these observations. We fit the peak emission in each time bin with a point source, and then computed the angular separation and position angle of the peak to the inferred position of the core from the radio astrometric measurements of \citet{atri2020radio}.
            
            While our time binning approach allows us to track the motion of components, it also leads to a decrease in sensitivity. Faint, fast moving sources that could not be detected in the full observation due to a large amount of smearing may not be detected in the short time bins. We therefore implemented a new technique to try to detect any components that might have been smeared below detectability due to large proper motions. 
        \subsubsection{Dynamic Phase Centre Tracking} \label{sec:shifting}
            In this technique, observations are split into a large number of discrete time bins, such that in each time bin the moving source travels across no more than 1/5 of the synthesized beam. With a user defined proper motion, the distance the source moves in each time bin is calculated, and the phase centre of the $uv$ data for the on-source observation in each time bin is shifted to account for this motion. The result is a series of time bins in which the moving peak appears in the same position relative to the updated phase centre. With the peaks of the moving source now aligned in each time bin, the $uv$ data from all of the time bins are concatenated, to produce an image whose phase centre tracks the component as it moves. This technique is distinct from the time binning described above, in that a single $uv$ data set and image are produced, rather than a series of individual, lower-sensitivity images. We implemented this technique using the ParselTongue Python interface to \textsc{AIPS} \citep{parseltongue}\footnote{Our implementation is available via a \href{https://github.com/Callan612/AIPS-Proper-Motion-Correction}{GitHub repository}}. This technique is similar to the synthetic tracking technique used to detect and track fast moving near-Earth asteroids and space debris at optical wavelengths \citep[e.g.][]{Tyson1992Kuiper,Yanagisawa2005,Shao2014Tracking,Zhai2014Tracking,Zhai2020Tracking}, although we implemented our technique in the visibility domain instead of in the image domain.
            
\section{Results}
\begin{figure}
        \centering
        \includegraphics[width=\linewidth]{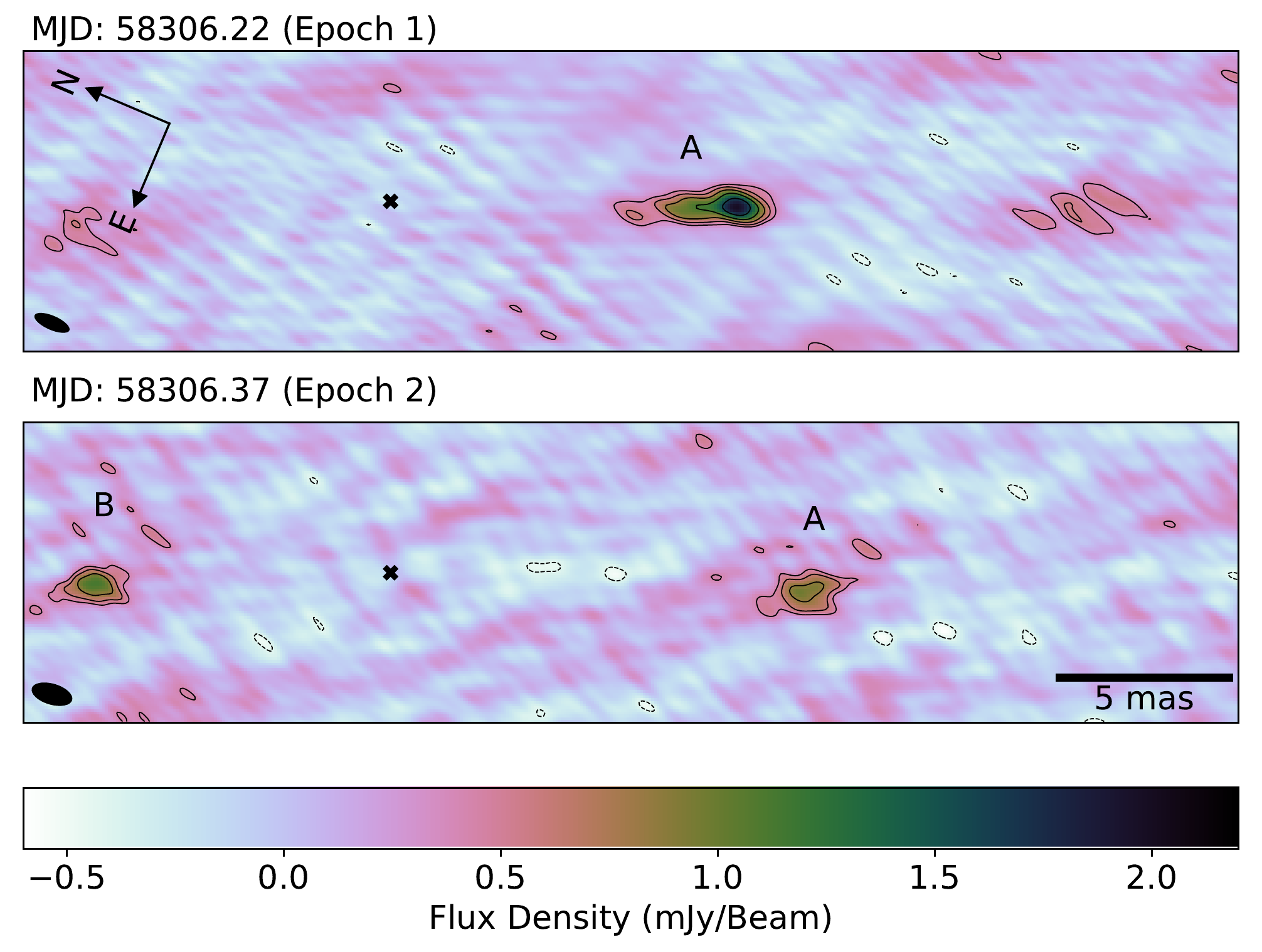}
        \caption{Images of MAXI J1820+070 from the first and second epoch. The contours mark $\pm\sigma\times (\sqrt{2})^n$ for $n=3,4,5,...$, where the rms noise $\sigma$ is $0.14$ mJy\,beam$^{-1}$ in the top panel and $0.15$ mJy\,beam$^{-1}$ in the bottom panel. The black crosses mark the inferred position of the core of J1820 \citep{atri2020radio}. The restoring beams for the images are $1.39$ mas x $0.52$ mas and $1.54$ mas x $0.77$ mas respectively, as marked by the black ellipses. Images have been rotated $67\degree$ counter-clockwise. The positions and flux densities of resolved components are summarised in Table~\ref{tab:components}. These images show the evolution of the approaching component (A) between the two epochs. A receding component (B) becomes visible in the second epoch.}
        \label{fig:obs1and2}
    \end{figure}

    First, we present the standard images of each epoch without time binning or dynamic phase centre tracking. Images of epochs 1 and 2 are shown in Fig.~\ref{fig:obs1and2}. The top panel shows the image of epoch 1. This image is dominated by a single elongated, asymmetric jet component, consisting of a compact bright knot, trailed by a diffuse tail that points towards the inferred core position of J1820 \citep{atri2020radio}. We refer to this as Component A. This component is $\approx4.5$ mas in length and has a total integrated flux density of $6.8\pm1.3$ mJy, where the uncertainty is $\sigma\sqrt{N_B}$ where $\sigma$ is the rms noise in the image and $N_B$ is the number of independent synthesised beams in the extended region as reported by the AIPS task TVSTAT. Its extended structure was initially attributed to smearing due to large proper motion, although this does not explain its asymmetric structure. 
        
    An image of the second epoch, made $3.5$ hours later, is shown in the bottom panel. The signal to noise ratio (SNR) in the second epoch is significantly lower than the first epoch as a result of the fainter emission of the components. Two ejecta are seen in this image; Component A to the south and Component B to the north. Component A is at a larger angular separation from the core position than in the first epoch, showing the motion of this ejection between the two observations. In this second epoch, Component A is no longer clearly extended along the jet axis, although it is not a perfect point source. The fainter emission from this component in this observation combined with the sparse {\it uv}-coverage means that we are not sensitive to any diffuse, extended structure. Component B is only significantly detected in the second epoch. J1820 was in the soft-intermediate state during both of these observations \citep{homan2020rapid} and so the core is not detected in our images, as expected \citep{fender2004towards}.
        
    For the fast moving ejecta, \citet{bright2020} determined that the component moving to the south is approaching, while the northern component is receding. This will be the same for the ejecta seen in these observations, i.e. Component A is approaching and Component B is receding. An image of the first epoch was originally presented by \citet{bright2020}, and Component A was identified. It was assumed that this component was the same as the approaching fast-moving, long-lived ejection seen travelling out to arcsecond-scale separations with eMERLIN, MeerKAT and the VLA. The VLBA detection of this component was used to constrain the motion of the fast moving ejecta by \citet{bright2020} and \citet{espinasse2020relativistic}.
    
    In our image of the first epoch, there appears to be a peak to the north of the core of J1820 at an angular separation of $11.4\pm0.1$ mas, which is in a similar position to where Component B is resolved in the second epoch. \citet{bright2020} identified this as the receding counterpart to Component A. This peak is not a compact point source, and only has a significance of $4\sigma$. This is comparable to other noise peaks elsewhere in the image, suggesting that caution is required in determining whether this is in fact a real detection.
    
    \subsection{Time Binning}
        Following the initial imaging of both epochs, we performed time binning (as per Section~\ref{sec:timebinning}) to determine the proper motion of Component A seen in Fig.~\ref{fig:obs1and2}. Component A was resolved in each time bin of the first epoch, and seen to be extended to a similar degree as in the image of the full observation (Fig.~\ref{fig:obs1and2}). Time binning did not reduce the smearing of this component, suggesting that motion within the observation is not likely the cause of its extended structure. We measured the position of Component A in each time-bin by fitting the peak with a point source in the image plane. The motion of the component appears to be far slower than that of the approaching fast moving ejection identified by \citet{bright2020}, with which it was initially identified. We fit the proper motion of this component with a constant velocity model. Extrapolating the motion of this component, we found that its position was consistent (within $2\sigma$) with two eMERLIN measurements made by \citet{bright2020} on MJD $58309.0$ and MJD $58310.03$. These two detections were originally considered to be anomalies that appeared alongside the approaching fast moving ejection. It was not clear if these detections were part of a larger structure of the fast moving ejection, the details of which had been resolved out, or if they were a separate ejection altogether. With only two measurements the motion of this component could not be adequately characterised, although it was estimated by \citet{bright2020} to have been launched at around the time of the fast moving ejecta based on its movement between the two epochs. The consistency of the eMERLIN measurements with the fit for the proper motion of Component A suggests they are the same ejection.
        
\begin{figure}
        \centering
        \includegraphics[width=\linewidth]{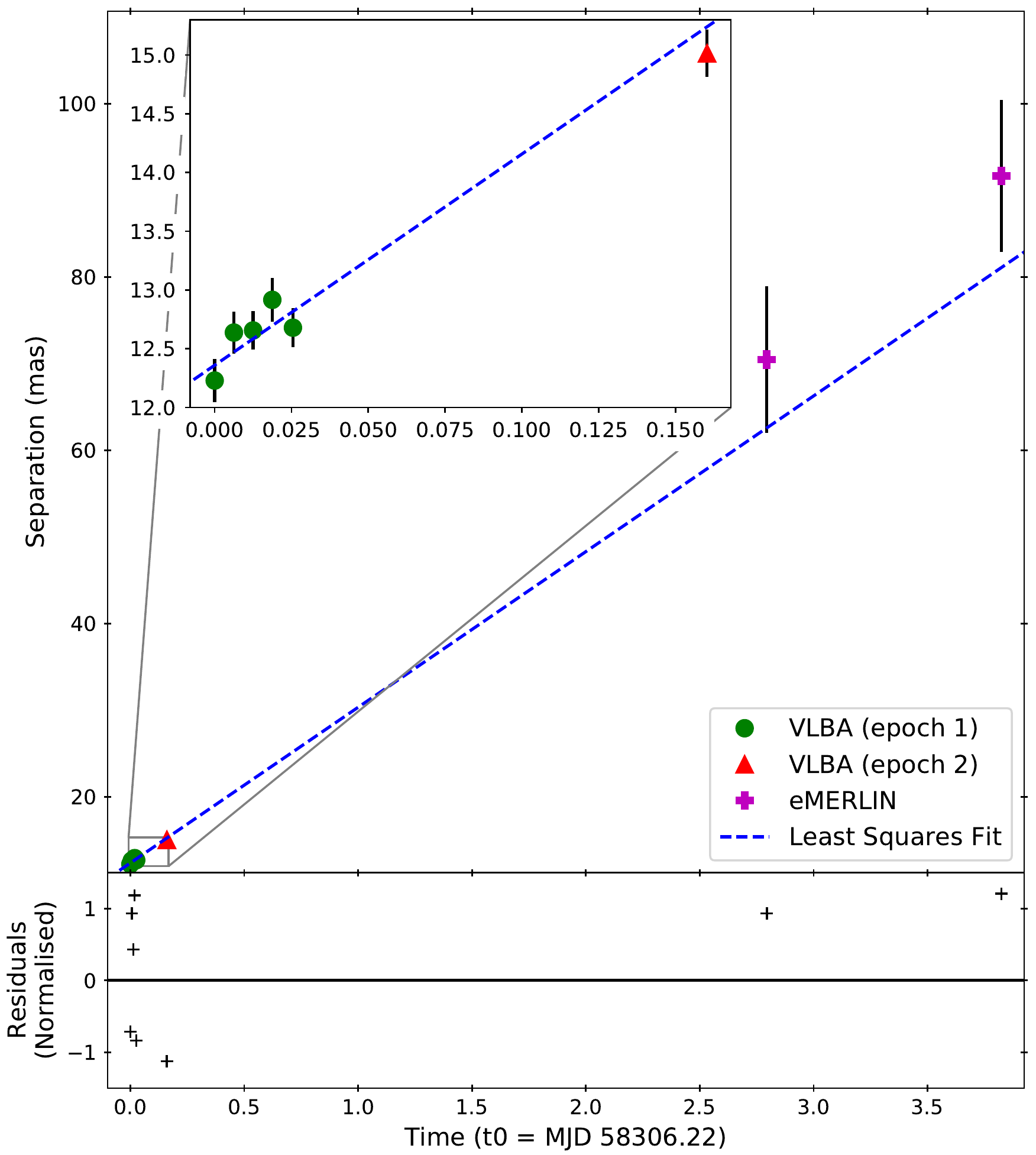}
         
        \caption{Fit for the proper motion of Component A seen in Fig.~\ref{fig:obs1and2}. The first panel shows the separation of the component from the inferred core position \citep{atri2020radio}, fit with a constant velocity model. The inset plot shows only the VLBA observations. $t_0$ here is the start-time of the first VLBA observation. The fit yields a proper motion $\mu_{\text{south}} = 18.0 \pm 1.1$ mas day$^{-1}$, and gives an ejection date of MJD $58305.60\pm0.04$. The bottom panel shows the residuals of the fit, calculated as the difference between the data and the model divided by the uncertainty. This component is distinct from the fast moving ejection described by \citet{bright2020} and \citet{espinasse2020relativistic}, travelling $\approx 5$ times slower and ejected $\approx 9$ hours earlier.}
        \label{fig:propermotion}
    \end{figure}

        Since there is no evidence of deceleration, we fit the proper motion of this component with a constant velocity model, using both the eMERLIN measurements and our VLBA measurements (using the time-binned data from epoch 1). Our best fit is shown in Fig.~\ref{fig:propermotion}, and yielded a proper motion of $\mu_{\text{south}} = 18.0 \pm 1.1$ mas day$^{-1}$ at a position angle of $-156.37 \pm 0.02\degree$ East of North. This gives an inferred ejection date of MJD $58305.60 \pm 0.04$. The uncertainties for the fitted parameters are given at the $1\sigma$ level. Fig.~\ref{fig:propermotion} shows some scatter in the separations of the measured peaks of the time-binned data, beyond what would be expected from the statistical uncertainties. Evolution of the extended structure of Component A during the observation could be responsible for this scatter in the position of the peaks. To account for this we added a systematic uncertainty of $0.13$ mas in quadrature to the statistical uncertainty found by fitting the peak positions. This systematic uncertainty was chosen in order to achieve a reduced $\chi^2$ value of 1 for the fit. This uncertainty corresponds to a quarter of the synthesized beam size, which is not unreasonable. Only the motion of Component A was fit, since Component B appears only in epoch 2, which could not be time binned due to lower SNR.

    \subsection{Dynamic Phase Centre Tracking}
\begin{figure}
        \centering
        \includegraphics[width=\linewidth]{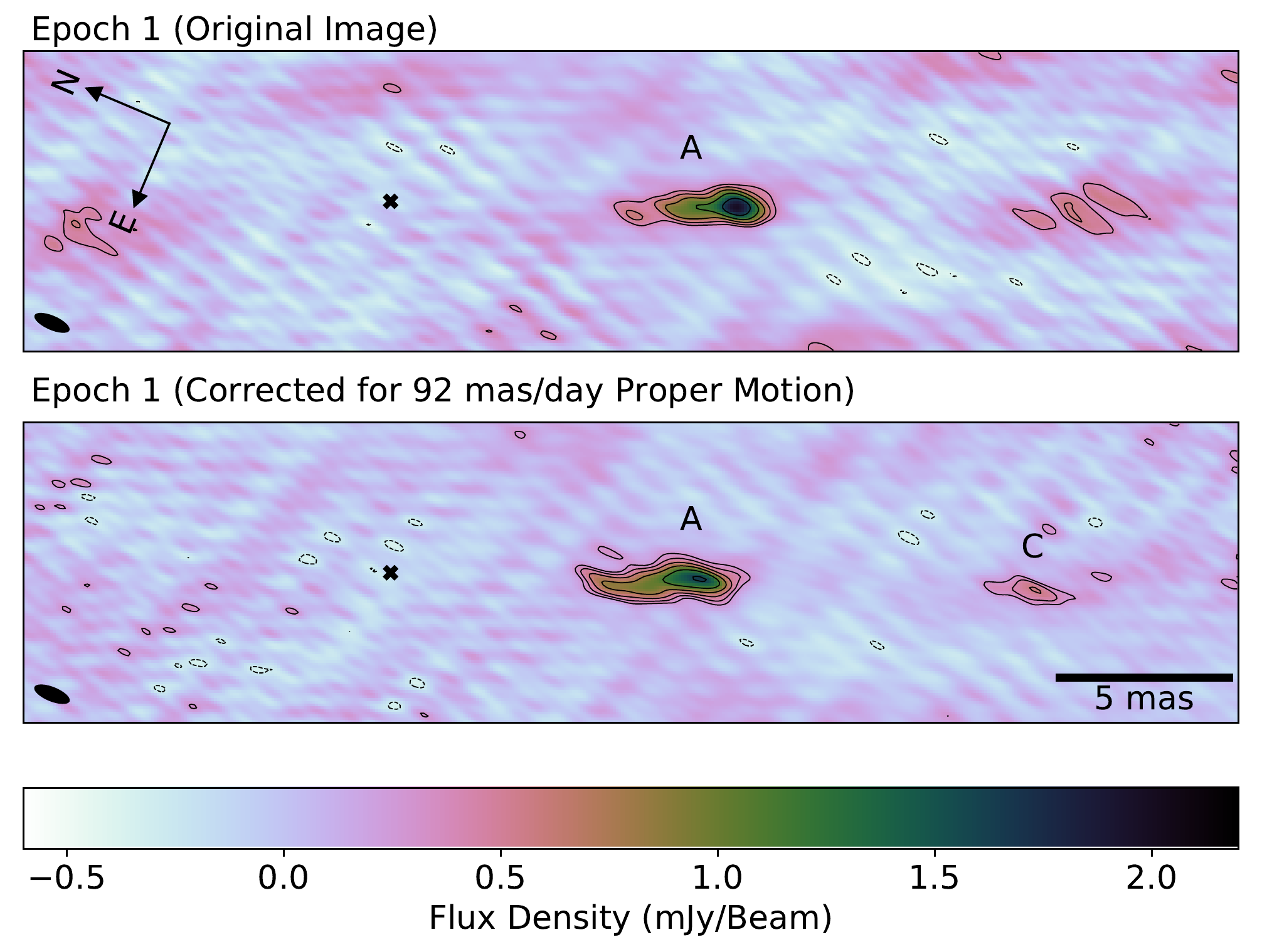}
         
        \caption{Two images of the first observation of MAXI J1820+070. The top panel shows an image made from the full observation. The second panel shows an image made following the technique described in Section~\ref{sec:shifting}, with the data split into 31 time bins, and those time bins phase shifted according to a proper motion of $92$ mas day$^{-1}$ at a position angle $-154.9\degree$ East of North \citep[as per the expected proper motion of the approaching fast moving component from][] {espinasse2020relativistic}. The contours mark $\pm\sigma\times (\sqrt{2})^n$ for $n=3,4,5,...$, where the rms noise $\sigma$ is $0.14$ mJy\,beam$^{-1}$ in the top panel and $0.11$ mJy\,beam$^{-1}$ in the bottom panel. The black crosses mark the inferred position of the core of J1820 \citep{atri2020radio} relative to the first time bin. The restoring beams for the images are $1.39$ mas x $0.52$ mas and $1.39$ mas x $0.53$ mas respectively, as marked by the black ellipses. Images have been rotated $67\degree$ counter-clockwise. The second image also reveals a $7\sigma$ detection of a fast travelling component (C), as described in Table~\ref{tab:components}. The RMS noise in these images is 0.14 and 0.11 mJy beam$^{-1}$, respectively. The newly detected component C was not detected in the original image due to smearing from its large proper motion, and is the same approaching fast travelling ejection described by \citet{bright2020} and \citet{espinasse2020relativistic}.}
        \label{fig:shift_92}
    \end{figure}

        With the identification of Component A as being distinct from the fast moving ejecta described by \citet{bright2020}, the absence of the approaching fast moving ejection is notable. The fit by \citet{espinasse2020relativistic} predicts the approaching fast moving component to be located at an angular separation of $24 \pm 7$ mas from the core in the first epoch, moving at a proper motion of $93.2\pm0.6$ mas day$^{-1}$. At this proper motion, the component should move 6 times the width of the synthesized beam during the first epoch, smearing its emission over that region. We therefore applied the dynamic phase centre tracking technique (as per Section~\ref{sec:shifting}) to try to detect this component. We applied this technique procedurally, stepping through a range of proper motions between $80-100$ mas day$^{-1}$ at the position angle fit by \citet{espinasse2020relativistic}. The dynamic phase centre tracking technique consistently revealed a component at an angular separation of $23.36 \pm 0.08$ mas from the core, at a position angle of $-157.9\pm 0.4\degree$ East of North, which we label Component C. Fig.~\ref{fig:shift_92} shows the original image of the first epoch in the top panel, and the image made when applying the technique for a proper motion of $92$ mas day$^{-1}$. At this proper motion Component C was brightest, detected with $7\sigma$ significance with a flux density of $0.82\pm0.11$ mJy\,beam$^{-1}$. Although applying the technique with a proper motion of $92$ mas day$^{-1}$ yielded the brightest detection, the component was detected at similar significance across the range of proper motions used, with a broad detection peak around $92$ mas day$^{-1}$. This component could not previously be robustly detected in the first epoch due to the smearing of the emission over multiple beams. In the original image of epoch 1 (Fig.~\ref{fig:obs1and2}) there appears to be some noise in the region where Component C was detected, likely due to smearing of the emission. Applying the dynamic phase centre tracking technique to the first epoch resulted in a decreased noise level from 0.14 to 0.11 mJy\,beam$^{-1}$. Applying the same technique to the second epoch with the same range of proper motions did not result in a detection, to a $5\sigma$ limit of $0.75$ mJy\,beam$^{-1}$.
        
        The dynamic phase centre tracking technique was also used to search for a receding component in both epochs for a range of proper motions between $10-50$ mas day$^{-1}$, however no new receding component was detected in either epoch. The $5\sigma$ detection limit is $0.7$ mJy\,beam$^{-1}$ for the first epoch and $0.75$ mJy\,beam$^{-1}$ for the second epoch. We also applied the dynamic phase centre tracking technique to the third epoch on July 8th, in search of a detection of any of Components A, B or C, but no detections were made above a 5$\sigma$ detection limit of $0.47$ mJy\,beam$^{-1}$.
    
        The angular separations, position angles, fitted peak flux densities and image noise levels for all our detected components are given in Table~\ref{tab:components}. 
        
    \begin{table*}
        \centering
        \caption{Fitted jet components from VLBA observations. We measured the positions of each component by fitting the peak of the emission with a point source, and we report the peak flux density. The uncertainty in the position and fitted flux density of components is the 1$\sigma$ statistical uncertainty as reported by the AIPS task JMFIT.}
        \label{tab:components}
        \begin{tabular}{|c|c|c|c|c|c|}
            \hline
            Epoch & Source                     & Separation    & Position Angle           & Fitted Flux Density             & Image RMS  \\ 
                  &                            & (mas)         & $\degree$ East of North  & (mJy\,beam$^{-1}$)                      & (mJy\,beam$^{-1}$) \\\hline \hline
            1     & Component A                & $12.6\pm0.1$  & $-156.5\pm0.4$           & \phantom{1}$2.68\pm0.14^a$      & $0.14$     \\ 
                  & \phantom{1}Component C$^b$ & $23.4\pm0.1^c$  & $-157.9\pm0.4$           & $0.82\pm0.11$                   & $0.11$     \\ \hline
            2     & Component A                & $15.0\pm0.1$  & $-158.7\pm0.3$           & $1.27\pm0.15$                   & $ 0.15$    \\ 
                  & Component B                & $10.9\pm0.1$  & $23.8\pm0.3$             & $1.34\pm0.15$                   &   \\ \hline
        \end{tabular}
        \begin{tabular}{l}
            $^a$ For the extended component we report on the fitted flux density of the peak of the emission and not the total integrated flux density.\\
            $^b$ Component detected following dynamic phase centre tracking.\\
            $^c$ Separation with respect to the core position in the central time bin.
        \end{tabular}
    \end{table*}
    
\section{Discussion}
    Three distinct components are resolved in these observations; Component A, seen approaching in both images in Fig.~\ref{fig:obs1and2}, Component B, seen receding in the second epoch (bottom image in Fig.~\ref{fig:obs1and2}), and Component C, detected in the first epoch by applying the dynamic phase centre tracking technique to correct for its large proper motion, seen in the bottom panel in Fig.~\ref{fig:shift_92}. Understanding each of these components contributes to the full picture of the ejection events in J1820. Each of these are now discussed in turn.
    
    \subsection{Component A}
        Component A appears to travel $\approx6$ times slower than the approaching fast-moving ejection described by \citet{bright2020} and \citet{espinasse2020relativistic}, which was detected as Component C in epoch 2. The two eMERLIN measurements made by \citet{bright2020} on MJD $58309.0$ and MJD $58310.03$ are consistent with the proper motion fit of Component A in our observations. With these eMERLIN detections alongside our time-binned VLBA measurements, we can properly characterise this component as a separate, slower-moving ejection with a proper motion of $18.0\pm1.1$ mas day$^{-1}$. Component A does not exhibit apparent superluminal motion, having an apparent velocity of $\approx0.31$c. The proper motion fits of Component C and its receding counterpart by \citet{bright2020} and \citet{espinasse2020relativistic} incorrectly included the position of Component A from the first VLBA data set. Following the identification of this component as being distinct from Component C, we revise these fits in Section~\ref{sec:fast component fit}.
        
        Component A is not travelling fast enough for smearing to account for the extended structure seen in the first image in Fig.~\ref{fig:obs1and2}, suggesting that the component is intrinsically extended. Furthermore, the component does not appear extended to as large a degree in the second epoch in Fig.~\ref{fig:obs1and2}. This is likely the result of sensitivity. By the second epoch the component has expanded, reducing its surface brightness as its emission has been spread over a larger area, making it harder to detect. The low SNR in this observation and the sparse {\it uv}-coverage then limits our ability to resolve the extended structure. LMXB jets from discrete ejection events are often modelled as point sources. However, extended ejecta have been observed before, as in the case of GRO J1655-40 \citep{hjellming_episodic_1995,tingay_relativistic_1995}, and could be due to a long-duration ejection event. Given the approximate size of Component A in the first epoch and its proper motion, then by assuming a steady and constant ejection velocity, we can estimate the duration of the ejection event to have been $\sim6$ hours. If Component A is expanding radially at a speed comparable to its bulk motion downstream, then the inferred ejection duration would be shorter than we estimate. Our data cannot constrain the expansion speed of Component A, however we do know that Component A cannot be expanding radially in a purely uniform way, given its elongated structure. Conversely if the initially ejected material is travelling slower than the later ejected material, or the working surface of the jet is significantly decelerated by its interactions with the interstellar medium (ISM) then the true ejection duration may be longer than we estimate, and hence we can only provide this rough estimation of the ejection duration by assuming constant velocity. As the ejection moves through the ISM, particles are shock-accelerated at the working surface, resulting in asymmetric emission. 
        
        Unlike the faster-moving Component C, we do not observe Component A to decelerate. However, since we can only track its motion out to $\sim 100$\,mas, we cannot place strong constraints on the absence of deceleration. The deceleration of Component C was attributed to continuous interaction with the ISM \citep{espinasse2020relativistic}. It has been suggested that X-ray binaries exist in low density bubbles \citep{Heinz2002,Hao2009Cavities}. A consequence of this is that ejecta would initially have ballistic motion before beginning to decelerate as they interact with the more dense ISM at the edge of the low density cavity. \citet{espinasse2020relativistic} suggested that this could be the case for Component C, although there was insufficient observational evidence to draw any strong conclusions. \citet{bright2020} also attributed the very slow decay rate of the radio emission from the jets to continuous interaction with the ISM. The constant velocity of Component A seen in Fig.~\ref{fig:propermotion} could argue against deceleration at small angular separations from the core, although this could alternatively be due to its relatively low proper motion, such that any small deceleration of Component A is not noticeable over its relatively short life-time.
        
    \subsection{Component B}\label{sec:componentB}
        Component B was seen receding in the second observation as shown in the second panel in Fig.~\ref{fig:obs1and2}. It is not immediately clear if this component is the counterpart to approaching Component A or C. The proper motion of Component A can be used to estimate the value of $\beta\cos{\theta}$ from 
        \begin{equation}
            \mu_{\text{app}}=\frac{\beta\sin{\theta}}{1-\beta\cos{\theta}}\frac{c}{d},
        \end{equation}
        where $\beta$ is the jet velocity normalised by the speed of light, $\theta$ is the inclination angle of the jet to the line of sight, and $d$ is the distance to J1820 \citep{mirabel1999}. \citet{atri2020radio} determined the jet inclination angle of the fast moving ejecta to be $(63\pm3)\degree$ using their measurement of the distance to J1820 and the proper motions of \citet{bright2020}. It is reasonable to assume \citep[in the absence of rapid precession as seen in V404 Cygni;][]{V404_Cygni} that in the time between the ejection of Components A and C the inclination angle has not changed significantly. Using the distance and inclination angle of \citet{atri2020radio}, we determine the value of $\beta\cos{\theta}$ for Component A to be $0.14\pm0.04$. At any given time, the ratio of angular separations of intrinsically symmetric approaching and receding ejecta is 
        \begin{equation}
            R = \frac{1+\beta\cos\theta}{1-\beta\cos\theta}.
        \end{equation}
        If we assume that Component B is the counterpart to Component A, the measured ratio $R$ would imply a value $\beta\cos{\theta}=0.16\pm0.02$. These two values of $\beta\cos{\theta}$ are consistent, suggesting Component B is likely the receding counterpart to the slow moving Component A. We still consider, however, the possible association of Component B with the fast moving Component C when we revise the fits of \citet{espinasse2020relativistic} in Section~\ref{sec:fast component fit}.
        
        If the noise peak seen in the first epoch in a similar position to Component B is a real detection, we cannot associate it with Component B, since it is slightly further from the core than Component B is in the second epoch. In Section~\ref{sec:fast component fit} we discuss whether this peak could be the receding counterpart to Component C.
                
       It is unclear why Component B is not detected in the first epoch. It is possible that the external (or internal) shocks that accelerate particles in the jet and generate emission had not yet occurred in Component B by the first epoch, possibly as a result of an anisotropy in the surrounding medium, or that Component B was still optically thick in the first epoch. It is also possible that Component B was obscured by some free-free absorbing medium during the first epoch, or that it was sufficiently extended in the first epoch such that its flux was spread over a number of beams and thus it could not be detected.
    
    \subsection{Component C}
        We applied our dynamic phase centre tracking technique to the first VLBA observation (epoch 1). When applying the technique for a proper motion of $92$ mas day$^{-1}$ we found a $7\sigma$ detection of a component, as shown in Fig.~\ref{fig:shift_92}. This component was previously undetected in this epoch due to smearing from its large proper motion. However, its proper motion and position lead us to identify it as the fast-moving component monitored by \citet{bright2020} and \citet{espinasse2020relativistic}. 
        
        From the estimate of the ejection size by \citet{bright2020} and assuming a constant expansion rate, this component would have been expanding at a rate of between $7$ and $187$ mas day$^{-1}$. This puts the size of Component C in the range of 2--40 mas in our first epoch, and 3--70 mas in our second epoch. The VLBA probes a maximum angular size of $\approx$7--10 mas, and so components larger than this would be resolved out. Component C was not seen in epoch 2, likely due to the fact that it has become too large and diffuse to be detected by the VLBA, even with dynamic phase centre tracking. It is important to note that the apparent expansion rate is not constant if the component is decelerating. The observed expansion speed of a component is modified by the relativistic Doppler factor $\delta=(\Gamma(1-\cos\theta))^{-1}$ \citep{Miller-Jones2006}, where $\Gamma=(1-\beta^2)^{-1/2}$ is the bulk Lorentz factor, and so as the component decelerates and $\Gamma$ decreases, the observed expansion rate increases. This is only important if the component is significantly relativistic, as is the case for Component C (see Section~\ref{sec:fast component fit}). However, our constraints on the expansion rate are not sufficient to constrain any decrease in Gamma.
    
    \subsection{Updated Proper Motion Fits}\label{sec:fast component fit}
    \begin{table*}
        \centering
        \caption{Fits for the proper motion of the fast moving ejecta to compare the inclusion of the receding VLBA component. All fits were made with eMERLIN, MeerKAT and VLA radio observations from \citet{bright2020}, and \textit{Chandra} X-ray observations from \citet{espinasse2020relativistic}. Fit (a) was made with both the approaching Component C and the receding Component B from the VLBA observations, and is shown with a dashed line in Fig.~\ref{fig:fast ejecta}. Fit (b) was made with only Component C, and is shown with a dash-dotted line in Fig.~\ref{fig:fast ejecta}. The quoted uncertainties in the fitted parameters are the 1$\sigma$ statistical uncertainties of the least squares fits.}\label{tab:fits}
        \begin{tabular}{|c|c|c|c|c|c|c|}
            \hline
            Fit & $\mu_{\text{south},0}$ & $\dot\mu_{\text{south}}$ & $\mu_{\text{north},0}$ & $\dot\mu_{\text{north}}$ & $t_0$             & $\chi^2$ \\ 
                & (mas day$^{-1}$)       & (mas day$^{-2}$)         & (mas day$^{-1}$)       & (mas day$^{-2}$)         & (MJD)             & (Reduced)\\ \hline \hline
            (a) & $88.8\pm2.6$           & $-0.31\pm0.04$           & $35.4\pm0.9$           & $-0.044\pm0.007$         & $58305.97\pm0.02$ & 3.0      \\ \hline
            (b) & $87.6\pm2.5$           & $-0.29\pm0.03$           & $35.9\pm0.8$           & $-0.048\pm0.007$         & $58305.95\pm0.02$ & 2.6      \\ \hline
        \end{tabular}
    \end{table*}
    \begin{figure}
        \centering
        \includegraphics[width=\linewidth]{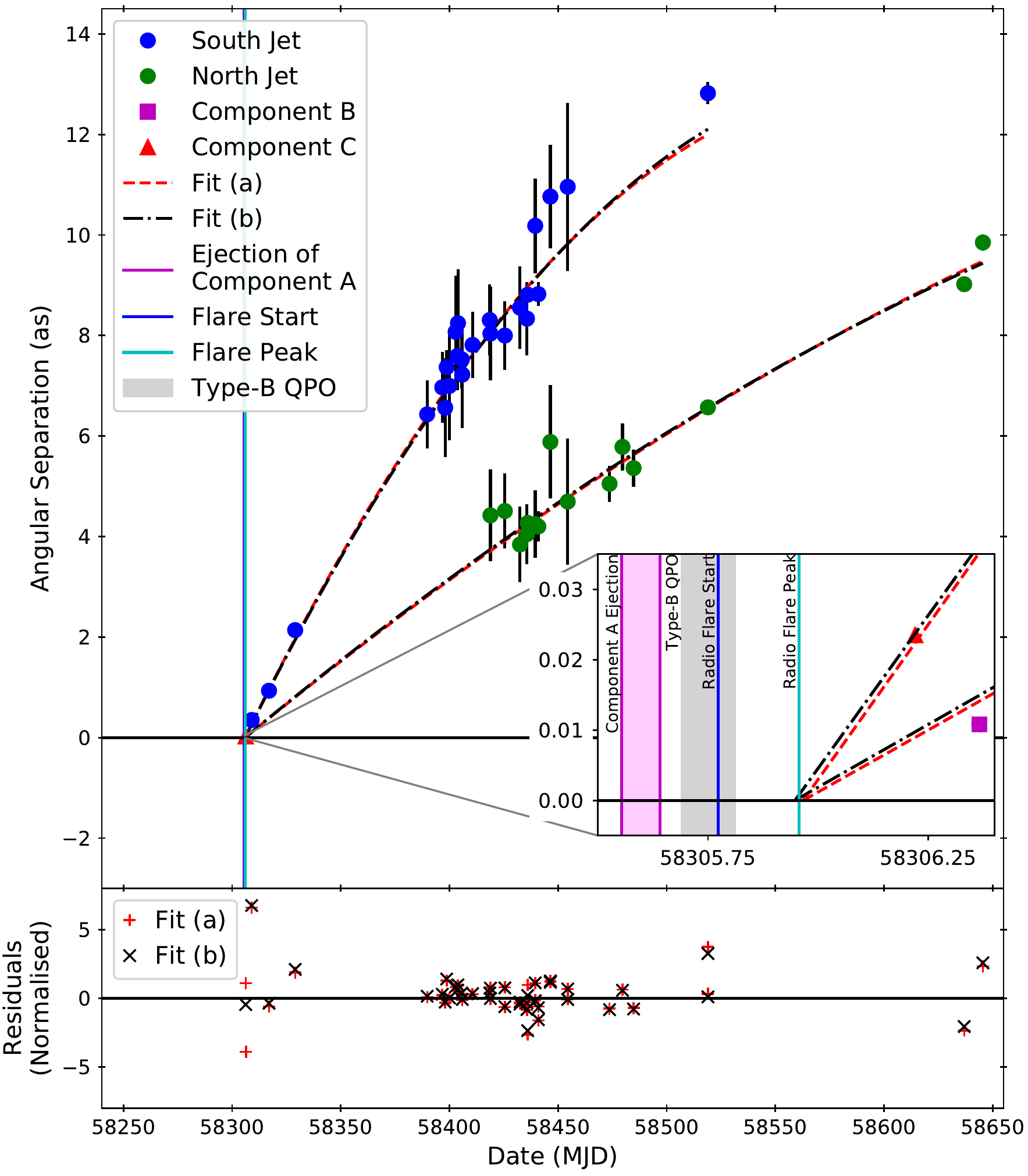}
         
        \caption{Revised constant deceleration fits for the proper motions of the fast travelling ejecta. The top panel marks the positions of the jets and the fits to their motion. Blue dots mark the approaching (southern) ejection. Green dots mark the receding (northern) ejection. The red triangle marks the approaching VLBA component detected using the dynamic phase centre tracking technique, labelled Component C in Fig.~\ref{fig:shift_92}. The magenta square marks the receding VLBA component from epoch 2, labelled Component B in Fig.~\ref{fig:obs1and2}. Fit (a) is the fit for the proper motion of these ejecta using both Components B and C. Fit (b) is the fit for the proper motion of these ejecta including Component C and excluding Component B. Fits are described in Table~\ref{tab:fits}. The shaded pink area shows the 1$\sigma$ bounds of the inferred ejection date of the slow moving Component A. The blue and red lines mark the start and peak of the AMI-LA radio flare, with the grey region marking the presence of the type-B QPO as reported by \citet{homan2020rapid}. The second panel shows the residuals to the two fits.}
        \label{fig:fast ejecta}
    \end{figure}
    
        The fit to the proper motion of the fast moving ejecta by \citet{espinasse2020relativistic} included Component A from our first VLBA epoch (Fig.~\ref{fig:obs1and2}). Following the identification of this as a distinct, slower-moving ejection, along with the new detection of Component C via dynamic phase centre tracking, we therefore revised this fit. We used the \citet{bright2020} radio data from eMERLIN, MeerKAT and the VLA, and the \citet{espinasse2020relativistic} X-ray data from \textit{Chandra} in our fits. We made two different fits, the first using both Components B and C, the second using only Component C, since we believe Component B may be the receding counterpart to Component A (Section~\ref{sec:componentB}). Both fits use a constant deceleration model as per \citet{espinasse2020relativistic}. These fits are shown in Fig.~\ref{fig:fast ejecta} and are outlined in Table~\ref{tab:fits}.
        
        The reduced $\chi^2$ value for the fit including both the VLBA components was $3.0$, and the value for the fit using only Component C was $2.6$. The decrease in the reduced $\chi^2$ value can be attributed to the low uncertainty in the position of Component B. These reduced $\chi^2$ values are smaller than for the fit made by \citet{espinasse2020relativistic} that included Component A, giving us confidence that they better represent the true proper motion of the fast moving ejecta (Component C).
        
        Our updated fits for the proper motion of the fast jet (which omitted Component A but included the newly detected Component C), did not significantly shift the inferred ejection date from that determined by \citet{espinasse2020relativistic}. However, the updated fits decrease the initial proper motion of Component C ($\mu_{0,\text{south}}$) from $\approx93$ mas day$^{-1}$ to $\approx88$ mas day$^{-1}$. The dynamic phase centre tracking technique yields the brightest detection of Component C at a proper motion of $92$ mas day$^{-1}$, however it is still detected at a similar significance at $88$ mas day$^{-1}$. The predicted position of the receding counterpart to Component C at the time of our second epoch disagrees with the measured position of Component B by 3 mas. While significantly larger than the synthesised beam, this is comparable to the size of Component B (Fig.~\ref{fig:obs1and2}), so the difference in position does not completely rule out an association. The fits for the proper motion of the fast moving ejecta with and without Component B are the same within uncertainty. Based solely on these fits, we could plausibly identify Component B as being the counterpart to either Components A or C. However, since Component B could plausibly be associated with Component A (Section~\ref{sec:componentB}), we conservatively choose to omit Component B in determining the proper motions of the fast moving ejecta. 
        
        Using these updated fits, we calculated the expected angular separation of the receding fast moving component (i.e. the receding counterpart to Component C) in the first epoch to be $10.8\pm0.8$ mas, which is consistent with the measured position of the small noise peak in the first epoch (Fig.~\ref{fig:obs1and2}). If this peak is a real detection then it may be the receding counterpart to Component C, although when we applied the dynamic phase centre tracking technique for the expected proper motion of this component, the peak disappeared. This suggests that this peak is unlikely to be the receding counterpart to Component C, and may therefore not be real emission.

        With the updated proper motion fits, the jet inclination angle $\theta$ from \citet{atri2020radio} can be updated. Assuming the jets are inherently symmetric, the jet inclination angle and jet velocities can be uniquely determined from
        \begin{equation}
            \tan{\theta}=\frac{2d}{c}\frac{\mu_{\text{app}}\mu_{\text{rec}}}{\mu_{\text{app}}-\mu_{\text{rec}}}
        \end{equation}
        \begin{equation}
            \beta\cos{\theta}=\frac{\mu_{\text{app}}-\mu_{\text{rec}}}{\mu_{\text{app}}+\mu_{\text{rec}}}
        \end{equation} \citep{Mirabel1994,fender1999}. Using the initial velocities of the updated fits, we calculate an inclination angle of $(64\pm5)\degree$, which is in agreement with the inclination angle found by \citet{atri2020radio} using the fits of \citet{bright2020}. With this updated inclination angle, the speed of the fast moving ejecta was calculated to be $0.97_{-0.09}^{+0.03}$c. From the proper motion of Component A, we used our revised inclination angle to determine its intrinsic speed to be $(0.30\pm0.05)$c. 
        
        Component C is travelling $\approx3.5$ times faster than Component A. Multiple ejection events where ejecta are travelling at similar velocities have previously been observed, such as with GRS 1915+105 \citep{fender1999,Dhawan_2000,MillerJones2005}. Multiple ejecta from the same system with significantly different velocities have also been observed, such as with the neutron star X-ray binary Scorpius X-1 \citep{Fomalont_2001}, and with the 2003 and 2009 outbursts of H1743-322 \citep{McClintock2009,millerjones2012}. The 2015 outburst of V404 Cygni showed multiple ejecta with different proper motions \citep{tetarenko2017,V404_Cygni}. It is not clear what sets the speeds of individual ejecta, and why they differ between ejection events, especially within the same outburst.
        
        The measured proper motions of intrinsically symmetric jets can be used to calculate a maximum possible distance to a source corresponding to $\beta=1$ \citep{mirabel1999},
        \begin{equation}
            d_{\text{max}}=\frac{c}{\sqrt{\mu_{\text{app}}\mu_{\text{rec}}}}.
        \end{equation}
        \citet{Fender2003ProperMotions} showed that close to this maximum distance the value of $\Gamma$ tends to infinity. Using the updated fits, we calculated a maximum distance of $3.11\pm0.06$ kpc. This is consistent with the distance measured by \citet{atri2020radio}, which means that we cannot place an upper limit on the value of $\Gamma$ for the fast moving ejecta. We do calculate a lower limit of $\Gamma>2.1$. For Component A we calculate $\Gamma=1.05\pm0.02$. 
    \subsection{Radio Flare}\label{sec:radio_flare}
\begin{figure}
        \centering
        \includegraphics[width=\linewidth]{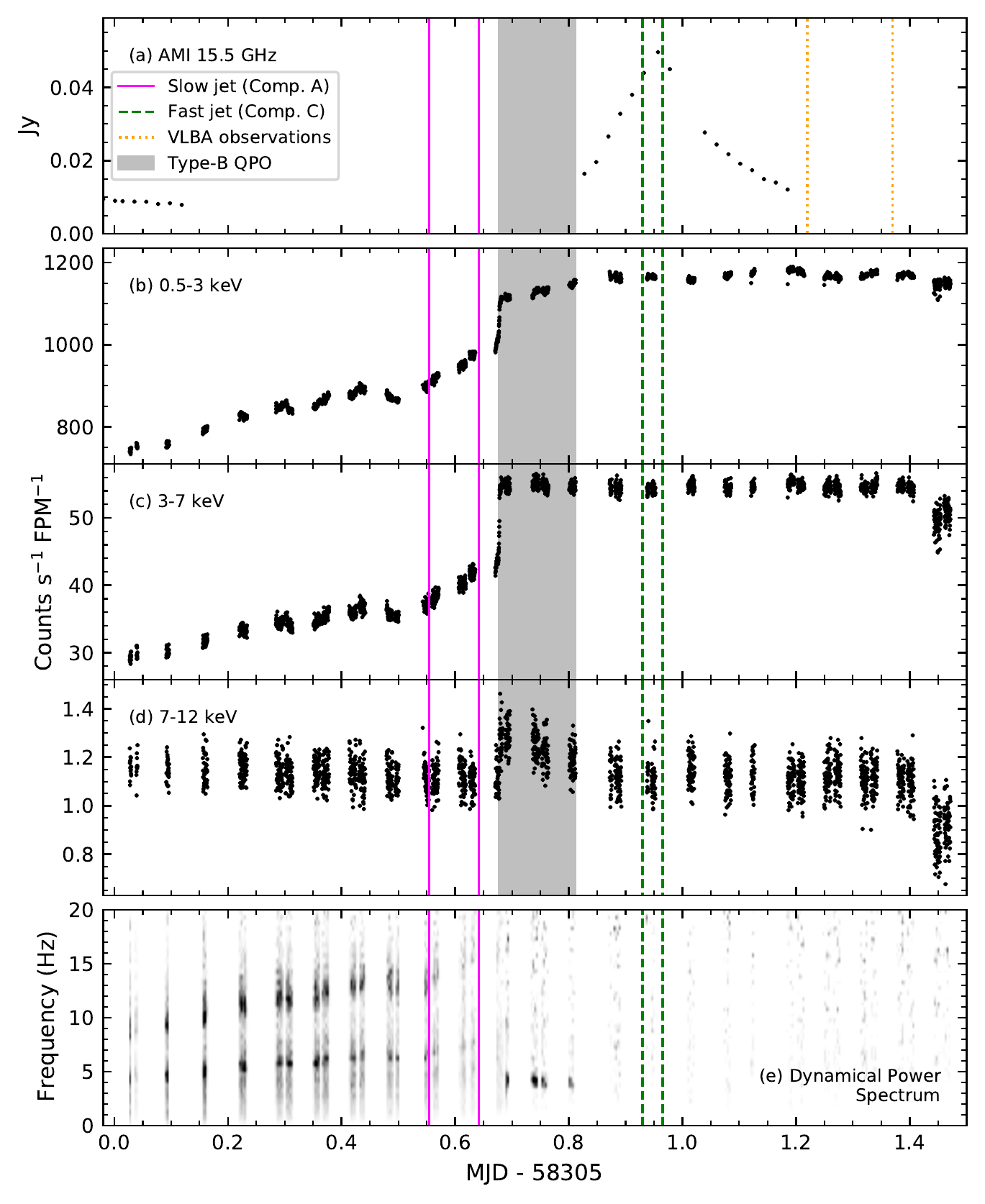}
         
        \caption{ AMI-LA radio and \textit{NICER} X-ray light curves of MAXI J1820+070 surrounding the inferred ejection dates of components A and C. Panel (a) shows the AMI-LA radio light curve at 15.5 GHz. Panels (b)-(d) show the X-ray count rates in the 0.5-3 keV, 3-7 keV, and 7-12 keV energy bands respectively. Panel (e) shows the 0.3-12 keV dynamical power spectrum \citep[data taken from][]{homan2020rapid}. The grey shaded region marks the presence of the type-B QPO and the green and pink lines mark the 1$\sigma$ bounds of the inferred ejection dates of components A and C respectively. The yellow lines in the first panel mark the observation dates of our first two VLBA epochs.}
        \label{fig:X-ray Timing}
    \end{figure}

        \citet{bright2020} and \citet{homan2020rapid} described a rapid radio flare that was contemporaneous with changes in the X-ray variability properties of J1820, and \citet{bright2020} associated the flare with the launch of the fast-moving ejecta (Component C). This radio flare is shown in Fig.~\ref{fig:X-ray Timing} alongside the X-ray light curves, the X-ray power density spectra and the inferred ejection dates of the slow and fast moving ejecta (Components A and C) from our proper motion fits. The first of our VLBA observations took place $\approx6$ hours after the peak of the radio flare. At this time, the interpolated flux density of the AMI-LA radio flare was $9.6$ mJy at 15.5\,GHz. In the first epoch Component A has a total integrated flux density of $6.8\pm1.3$ mJy, suggesting that this component is primarily responsible for the radio flare. This is consistent with our proper motion constraints, which imply that Component C was ejected contemporaneously with the peak of the flare, such that it could not have been responsible for the rise phase.
        
        Furthermore, our derived jet parameters imply that Component C and its counterpart will be significantly Doppler-deboosted, reducing their contribution to the total flux density. For intrinsically symmetric jets, the ratios of the received flux density from the approaching and receding components ($S_{\rm a}$ and $S_{\rm r}$ respectively) to the emitted flux density in the rest frame of the source ($S_0$) are given by 
        \begin{equation}\label{eq:doppler}
            \frac{S_{\rm a,r}}{S_0} = \left(\frac{1}{\Gamma\left(1\mp\beta\cos\theta\right)}\right)^{k-\alpha},        \end{equation}
        where $\alpha$ is the spectral index of the emission ($S_\nu\propto\nu^\alpha$) and $k$ describes the geometry of the ejecta \citep{mirabel1999}. In this case $\alpha=-0.7$ for optically thin synchrotron emission, and $k=3$ for discrete ejecta. For the fast moving ejecta, this Doppler boosting factor is $<0.39$ for the approaching component (Component C) and $<0.02$ for the receding component. Upper limits were calculated using the lower limit of $\beta$. While we do not know the intrinsic flux density of Component C, this could explain a reduced contribution to the AMI-LA radio flare. This also suggests that the receding counterpart to Component C could be significantly de-boosted below the detection threshold, and so the receding Component B detected in the second epoch likely corresponds to the approaching Component A. It is important to note that the emitted flux density $S_0$ should vary over time as the ejecta expand and fade, so the flux densities of approaching and receding components can only be directly compared using Equation~(\ref{eq:doppler}) when they are at equal angular separation from the core \citep{Miller-Jones2004JetEvolution}. Using Equation (\ref{eq:doppler}) and the integrated flux density of Component A in the first epoch, we calculated the expected integrated flux density of its symmetrically receding counterpart to be $2.6\pm0.6$ mJy at an angular separation of $12.6\pm0.1$\,mas from the core. Component B does not reach this angular separation until after the second epoch, so should have been brighter in epoch 2. However, this is the integrated flux density spread over several beams, making our lower measured peak flux density consistent with this prediction.
        
        \citet{bright2020} and \citet{homan2020rapid} fit an exponential decay to the AMI-LA radio light curve underlying the rise and peak of the radio flare, which was attributed to the quenching of the radio core, from which the flare appears to be distinct. X-ray observations of J1820 suggest that the system was in the soft-intermediate state during the rise and peak of the AMI-LA radio flare \citep{homan2020rapid}, and hence we don’t expect the core to be bright and contributing significantly to the AMI-LA radio flux for a substantial duration before the VLBA observations in which it was undetected.
        
        By assuming that the peak of the radio flare corresponds to the point at which the synchrotron emission of the jet becomes optically thin, \citet{bright2020} used the peak flux of the radio flare to estimate the internal energy of the jet knot that corresponds to the radio flare to be $E_i = 2\times10^{37}$ erg. As per \citet{fender2019synchrotron}, using the peak flux density of the radio flare ($\sim46$ mJy) at 15.5 GHz at a distance of 2.96 kpc, we estimate a minimum energy magnetic field strength for Component A of $\sim2.6$ G. This is of a similar order of magnitude to the minimum magnetic field strengths calculated from radio flares in V404 Cygni, Cygnus X-3 and GRS 1915+105 \citep{fender2019synchrotron}.
        
    \subsection{Ejection Events}\label{sec:ejection event}
        The ejection date of Component A is $4\pm1$ hours before the beginning of the radio flare and $2\pm1$ hours before the beginning of the type-B QPO period and the associated rise in the soft X-ray count rate as shown in Fig.~\ref{fig:X-ray Timing}. Given its asymmetric structure in the first epoch, we measure the proper motion of the leading peak of Component A, and so the inferred ejection date from the proper motion fit only marks the beginning of the ejection of this component. Since the estimated ejection duration of Component A is $\sim6$ hours, the extended tail of Component A would have been ejected during the type-B QPO period, and during the rise of the radio flare. This would provide some of the strongest evidence to date for jet ejection contemporaneous with a specific X-ray timing signature from the accretion flow.
        
        Alternatively, given its proximity to the beginning of the type-B QPO period, we could consider a scenario where the beginning of the ejection of Component A coincides with the beginning of the type-B QPO period. For this to be possible the component must undergo rapid deceleration to reach the separation and constant velocity with which it is seen travelling in the first epoch. In this scenario Component A is ejected at the beginning of the Type-B QPO period and decelerates until it is travelling at a velocity of $18.0$ mas day$^{-1}$ at a separation of $12.2$ mas at the beginning of the first epoch. Enforcing these conditions we calculated a lower bound on the initial velocity and acceleration of $28.7$ mas day$^{-1}$ and $-20.2$ mas day$^{-2}$ respectively.
        
        As discussed in Section~\ref{sec:radio_flare}, the integrated flux density of Component A, the peak flux density of Component C and the high level of deboosting of Component C's undetected receding counterpart suggest that the radio flare is due to the slow moving Component A. The time delay relative to its inferred ejection date could be due to optical depth effects \citep[e.g.][]{tetarenko2018}. When Component A is ejected it is initially optically thick. As the ejection expands adiabatically it becomes optically thin, and the 15.5 GHz emission probed by AMI-LA peaks and then decreases as the ejection expands further. The other possible explanation would invoke the shock-in-jet model \citep[e.g.][]{jamil2010,Malzac2014}, which posits that the flare is the result of internal shocks when a shell of ejected material collides with previously-ejected, slower-moving material. The delay between the ejection date and the radio flare is due to the time it takes for the ejection to travel out to the distance at which these shocks takes place. No second radio flare corresponding to the ejection of the fast moving ejecta was observed, possibly because the delays discussed above could have led it to occur during a gap in the AMI-LA observations.
        
        The ejection dates of Components A and C suggest that prior to the first VLBA epoch, Component C must have either passed by or collided with Component A in order for it to be found at a larger angular separation from the core. Based on our fits for these components, the calculated intersection time of these two ejections is MJD $58306.06\pm0.03$. This occurs $\approx 3.8$ hours before our epoch 1. There is no evidence of an interaction between the two ejecta in the AMI-LA radio and \textit{NICER} X-ray light curves \citep{homan2020rapid}. Component C may have been ejected at a slightly different angle to Component A, as a result of a small precession of the accretion disk about the jet axis, as was seen in GRO 1655-40 \citep{hjellming_episodic_1995} and to a much larger extent in V404 Cygni \citep{V404_Cygni}. Observations of J1820 in the hard state prior to its July 2018 outburst with the Hard X-ray Modulation Telescope (\textit{Insight}-HXMT) revealed Low Frequency QPOs across a range of energies. It was suggested that the observed QPO behaviour likely resulted from precession of the base of the compact jet \citep{Ma2021}. If the jet (or the accretion disk) is indeed precessing and the two components are launched at slightly different angles, their interaction may have been minimal.
  
        Despite the inferred ejection date of Component C and its receding counterpart being $\approx7$ hours after the beginning of the type-B QPO period, the evolution of the X-ray properties could still be linked to the ejection of these fast moving ejecta if they took time to accelerate up to the initial velocity with which they were observed. An assumption of the constant deceleration model is that the ejecta are launched at time $t_0$ with an initial proper motion $\mu_0$, and does not account for any initial acceleration period, which would move the ejection date earlier. In this case it could have been ejected alongside Component A during the type-B QPO period.  However, we have no empirical evidence for the prolonged period of acceleration that would be required. Alternatively, it could take time for the accretion flow changes observed as the X-ray flare and the change in the timing properties to manifest themselves in the ejection of this second, faster jet component.
        
        \citet{millerjones2012} and \citet{Russell_2019} attempted to identify a connection between the switch from type-C to type-B QPOs and the ejection of discrete jets in H1743-322 and MAXI J1535-571 respectively. While their data  were suggestive of a connection, due to gaps in their radio and X-ray coverage they could not conclusively associate these phenomena. They also reported on a rise in the soft X-ray count similar to what is seen here, although the rise in soft X-ray count occurred prior to the QPO switch, unlike what is seen here. The uncertainty on the ejection dates of H1743-322 and MAXI J1535-571 were $\sim0.5$ days and $\sim2$ days respectively. As a result of higher-cadence VLBA observations, time binning, and the strong lever arm of the downstream eMERLIN observations we have been able to constrain the ejection time to within an hour. This, in combination with dense NICER X-ray coverage has allowed for the association of the ejection of jet material and a change in the X-ray timing properties. Based on geometrical arguments (e.g. the dependence of QPO strength on the inclination angle) \citet{Motta2015Geometrical} suggested that type-B QPOs are related to jets, and although the launching of discrete ejecta have been seen at similar times to type-B QPOs, this is the first time an ejection has been shown to be  occurring during the emergence of type-B QPOs.
        
        The physics underlying the disk/jet connection is an area of active investigation, and as reviewed in, e.g., \citet{Ingram2020}, the origin of variable X-ray signatures such as QPOs is not yet well determined. 
        Numerical simulations of black hole accretion have not yet fully captured state transitions, but they already offer some interesting considerations that could guide future campaigns and the interpretation of phenomena such as the ejecta we describe. During accretion, the disk carries in and/or generates magnetic fields that can eventually saturate near the event horizon and provide enough pressure to disrupt the inflow (magnetically arrested disks, or MAD; \citealt{Igumenshchev2003,Igumenshchevv2008,Tchekhovskoy2011}). As the magnetisation at the horizon is directly linked to jet power \citep[e.g.][]{Komissarov2007,Tchekhovskoy2011}, one could expect jet ejecta launched during MAD states to be faster. The launching of the faster Component C subsequent to the slower Component A during a state transition is qualitatively consistent with the gradual build up of magnetic flux as the source brightens in outburst. The association with type-B QPOs is less clear, however magnetic reconnection in MAD simulations can drive ejecta and changes in variability \citep{Dexter2014}. While this effect has so far mostly been explored for flares in Sgr A* \citep[][Chatterjee et al., subm.]{Dexter2020,Porth2021}, higher resolution simulations are starting to reveal more significant dynamical changes \citep{Ripperda2020}, allowing the exploration of links between MAD-induced variability and type-B QPOs, with ejecta during state transitions.
        
        In summary, we have shown for the first time that an ejection event was occurring during the transition from type-C to type-B QPOs. This ejection (Component A) appears to be responsible for the subsequent radio flare. It is not clear if the faster moving Component C is also linked to the change in X-ray count rate and timing properties via some delayed ejection mechanism or acceleration period. The delay between the accretion flow evolution and the fitted ejection time of Component C means that we cannot identify a definitive signature of ejection for this component, should one exist.

\section{Conclusions}
    We provide detailed analysis of two VLBA observations of MAXI J1820+070 during the hard to soft state transition on MJD $58306$, and identify an approaching slow moving ejection, only seen previously in two eMERLIN observations but erroneously associated to a faster moving ejection in previous works. Via the time binning of these VLBA observations, the proper motion of this ejection was determined to be $\mu_{\text{south}} = 18.0 \pm 1.1$ mas day$^{-1}$ with an inferred ejection date of MJD $58305.60 \pm 0.04$. This ejection of the slow moving component began $4.2$ hours before the beginning of the rise of the radio flare and $2$ hours before the beginning of the type-B QPO period. The ejection of this component lasted for $\sim6$ hours and thus was contemporaneous with the changes in X-ray count rate and timing properties and the rise time of the radio flare. 
    
    A new technique was implemented to mitigate the effects of smearing in images due to large proper motions, which resulted in the $7\sigma$ detection of the approaching fast moving ejection in a VLBA observation in which it was previously undetected. Following this, the fits to the proper motion of the fast moving ejecta were updated, yielding an ejection date of MJD $58305.97\pm0.02$, which corresponded to the peak of the radio flare. We used these revised fits to calculate a jet inclination angle of $(64\pm5)\degree$, and jet velocities of $0.97_{-0.09}^{+0.03}c$ for the fast moving ejecta ($\Gamma>2.1$), and $(0.30\pm0.05)c$ for the newly identified slow moving ejection ($\Gamma=1.05\pm0.02$). It is unclear what is responsible for the large difference in velocities of these ejecta. We have shown that the approaching slow moving component is responsible for the radio flare, and is likely linked to the the switch from type-C to type-B QPOs, while no definitive signature of ejection was identified for the fast moving component.
    
\section*{Acknowledgements}
The National Radio Astronomy Observatory is a facility of the National Science Foundation operated under cooperative agreement by Associated Universities, Inc.  This work made use of the Swinburne University of Technology software correlator, developed as part of the Australian Major National Research Facilities Programme and operated under licence. SM is thankful for support from an NWO (Dutch Research Council)  VICI  award,  grant  Nr.  639.043.513. TMB and TDR acknowledge financial contribution from the agreement ASI-INAF n.2017-14-H.0. VT is supported by programme Laplas VI of the Romanian National Authority for Scientific Research. The authors also wish to recognize and acknowledge the very significant cultural role and reverence that the summit of Maunakea has always had within the indigenous Hawaiian community. We are most fortunate to have the opportunity to conduct observations from this mountain.

\section*{Data Availability}
The VLBA data are publicly available from the \href{https://archive.nrao.edu/archive/}{NRAO archive}, under project code BM467. The dynamic phase centre tracking script is available via a  \href{https://github.com/Callan612/AIPS-Proper-Motion-Correction}{GitHub repository}.



\bibliographystyle{mnras}
\bibliography{mj1820} 




\appendix



\bsp	
\label{lastpage}
\end{document}